\documentclass[review]{elsarticle}

\usepackage{amssymb}

\newcommand{\clR}{\mathcal{R}}

\newcommand{\clC}{{\cal C}}

\newcommand{\hL}{\hat{\cal L}}
\newcommand{\hF}{\hat{\cal F}}

\newcommand{\clK}{{\cal K}}
\newcommand{\clN}{{\cal N}}
\newcommand{\prt}{\partial}

\newcommand{\frD}{D_{\rm fr}}

\newcommand{\rgl}{\rangle}
\newcommand{\lgl}{\langle}

\newcommand{\be}{\begin{equation}}
\newcommand{\ee}{\end{equation}}
\newcommand{\bea}{\begin{eqnarray}}
\newcommand{\eea}{\end{eqnarray}}

\begin{document}

\begin{frontmatter}

\title{Maximum therapeutic effect of glioma treatment by radio-frequency electric field}

\author{Alexander Iomin}
\ead{iomin@physics.technion.ac.il}


\address{Department of Physics, Technion, Haifa, 32000, Israel}

\address{Max-Planck-Institute for the Physics of Complex Systems, Dresden,
Germany }

\begin{abstract}
The influence of a radio-frequency electric field on glioma brain
cancer development is considered. Specifically, the effectiveness
of this medical technology against invasive cells with a high
motility, when switching between migrating and proliferating
phenotypes takes place, is addressed. It is shown that glioma
development under the external field treatment can be modelled in
the framework of the continuous time random walk, where the
segregation of proliferating from not proliferating cells by the
external electric field is naturally accounted for. The
constructed reaction-transport equation is based on the fact that
cytokinesis affects the cell dynamics strongly, and the cell
kinetics is non-Markovian, which is reflected in both the
transport and reaction properties of cells. As a result of the
interplay between cell proliferation and cell degradation,
migrating cancer cells spread freely without being influenced by
the details of the degradation of the proliferating cells, which
leads to a decrease in the effectiveness of the treatment.

\end{abstract}


\end{frontmatter}

\section{Introduction}\label{sec:intro}

A new therapeutic method suggested in \cite{palti1,palti2,palti3}
for non-invasive treatment of glioma brain cancer by a
radio-frequency electric field also opens new directions of
understanding of glioma development. A specific question that
emerges is the effectiveness of this new medical technology
against invasive cells with high motility, when switching between
migrating and proliferating phenotypes takes place. As is well
known, one of the main features of malignant brain cancer is the
ability of tumor cells originating in the glioma to invade normal
tissue at a distance from the multi-cell tumor core. This abnormal
motility constitutes the metastasis feature of brain cancer,
causing treatment failure \cite{Geho}. This problem relates to
modelling of the dynamics of cancer glial cells in heterogeneous
media (as brain cancer is) in the presence of a radio-frequency
electric field, which acts as a tumor treating field (TTField)
\cite{palti1,palti2,palti3}. As reported, this transcranial
treatment by a low-intensity (1-3 V/cm), intermediate-frequency
(100-200 kHz) alternating electric field, produced by electrode
arrays applied to the scalp, destroys cancer cells that are
undergoing division, while normal tissue cells are relatively
unaffected. An important result of this new treatment technology
is that the survival period is increased twofold.

Glioma is one of the most recalcitrant brain diseases, with a
standard treatment survival period of 7-12 months
\cite{palti3,stupp}. One of the main mechanisms of this
devastating manifestation is the migration-proliferation dichotomy
of cancer cells. This phenomenon was first observed in clinical
investigations \cite{Giese1,Giese2}, where it was shown that
cancer cells in the outer invasive zone of glioma possess the high
motility property, while the proliferation rate of these migratory
cells is essentially lower than those in the tumor core. This
anti-correlation between proliferation and migration of cancer
cells, also known as the "Go or Grow" hypothesis (see discussions
in \cite{garay,Corcoran}), suggests that cell division and cell
migration are temporally exclusive phenotypes \cite{Giese1}. The
phenomenon that tumor cells defer proliferation for cell migration
was also experimentally demonstrated in
\cite{garay18,garay26,garay37}\footnote{This kind of
migration-proliferation dichotomy was also found at metastatic
behavior of breast cancer \cite{breastcancer}.}. The switching
process between these two phenotypes is still not well understood.
Moreover, it should be mentioned that conflicting data appear in
the literature concerning the Go or Grow hypothesis; details of
discussions on this can be found in \cite{Corcoran,garay}.

Extensive theoretical modelling followed these findings, and much
effort is being invested in developing relevant models with
relevant switching mechanisms of the glioma cells to understand
this dichotomy, when the high motility suppresses cell
proliferation in the outer invasive region, while the immobile
cancer cells have a high proliferation rate like the cells in the
cancer core. There are different approaches that have resulted in
several phenomenological models, which have supplied also
experimental \textit{in vitro} confirmation data
\cite{Khain,Sander}. Comprehensive discussions of these models
appear in \cite{Khain,Deutsch,Chauviere,kolobov,fir2011}.

The main subject of the paper is the fractional kinetics of the
new therapeutic technology \cite{palti1,palti2,palti3}, where the
TTField acts against invasive proliferating cancer cells. The
TTField segregates proliferating cells (either transporting or
not) from transporting cells out of cytokinesis. As stated in Ref.
\cite{garay}, on the individual cell level, the processes of
migration and proliferation are separated in time. The TTField
``fills'' this dichotomy of a single cell. These facts are
important for a continuous time random walk (CTRW) description,
since it is based on the dynamics of a single particle ({\it
e.g.}, see review \cite{klafter}). Therefore, we follow the CTRW
consideration presented in \cite{iom2006}, where a mechanism for
entrapping transporting cells by fission was suggested, and the
migration-proliferation dichotomy appears naturally. Therefore,
the efficiency of the TTField action can be evaluated in the
framework of the fractional kinetics of glioma development under
the TTField treatment. The first study of this kinetics in the
framework of a fractional comb model \cite{iom2012} showed that
the TTField efficiency is restricted by its space dispersion. In
particular, it depends on the fractal mass dimension of the cancer
cells. In the present research, we propose a general,
probabilistic consideration of cancer cell kinetics and obtain a
solution for an arbitrary distribution of proliferating cells,
which can change in time and space. In the framework of the
constructed model, we show that the glioma development corresponds
to the \textit{compensated} cancer development, when the TTField
compensates cell proliferation, and we consider this solution as
the maximum therapeutic effect of glioma treatment due to the
TTField.

\section{Time multi-scaling of the CTRW modelling
of cell transport}\label{sec:ctrw1}

\subsection{Fission and motility times}
A simplified scheme of the migration-proliferation dichotomy of
cancer cells can be considered by virtue of two time scales of
tumor development \cite{iom2006}. The first corresponds to the
biological process of cell fission with random duration ${\cal
T}_f$. The second process is cell transport with random duration
${\cal T}_t$. During the time scale ${\cal T}_f $, the cells
interact strongly with the environment, and the motility of the
cells is vanishingly small.
During the second time ${\cal T}_t$, interaction between the cells
is weak and the motility of the cells leads to cell invasion,
which is a very complex process controlled by matrix adhesion
\cite{Geho,dongen,brabek}. It involves several steps, including
receptor-mediated adhesion of cells to the extracellular matrix
(ECM), matrix degradation by tumor-secreted proteases
(proteolysis), detachment from ECM adhesion sites, and active
invasion into the intercellular space created by protease
degradation. It is convenient to introduce a ``jump'' length,
$X_t$, of these detachments as the distance that a cell travels
during the time ${\cal T}_t$. Hence, the cells form an initial
packet of free spreading particles, and the contribution of cell
dissemination to the tumor development process consists of the
following time sequences: ${\cal T}_f(1){\cal T}_t(2){\cal
T}_f(3)$. There are different realizations of this chain of times,
due to different durations of ${\cal T}_f(i)$ and ${\cal T}_t(i)$,
where $i=1, 2, \dots$. Therefore, one concludes that transport is
characterized by random values ${\cal T}(i)$, which are waiting
(or self--entrapping) times between any two successive jumps of
random length $X(i)$. This phenomenon is known as a continuous
time random walk (CTRW) \cite{MW}. It arises as a result of a
sequence of independent identically distributed random waiting
times ${\cal T}(i)$, each having the same pdf $\psi(t), ~t>0$ with
a mean characteristic time $T$, and a sequence of independent
identically distributed random jumps, $x=X(i)$, each having the
same pdf $p(x)$ with a jump length variance $\sigma^2$, equal to
the square averaged intercellular/cellullar space $\sigma\sim
10^{-3}{\rm cm}$. It is worth mentioning that a cell ``carries''
its own trap, by which it is set apart from the transport process.
This process of self--entrapping differs from the standard CTRW,
where traps are external with respect to the transporting
particles. The probability to ``escape from $i$-th trap'' of
fission is locally determined by the Poisson distribution
$e^{-t/\tau_i}$ with different time scales, or even random scales.

\subsection{Example of power law: a toy model}

This absence of a unique time scale leads to the power law
distribution of fission times. For example, one can suppose that
self-entrapping for different generations of cells has different
mean characteristic time scales. We consider that the $j$-th
generation of self-entrapping is the Poisson process
$$w_j(t)=\tau_j^{-1}\exp\left(-t/\tau_j\right)$$
with the characteristic time scale $\tau_j=\tau^j$, where
$\tau_1=\tau$ is now an average cell division time for the first
generation. Therefore, following \cite{shlezinger1,blumen} and
repeating exactly the analysis of Ref. \cite{blumen}, we obtain,
by taking into account events occurring on all time scales, the
distribution
$$\psi(t)=\frac{1-b}{b}\sum_{j=1}^{\infty}b^{j}\tau^{-j}
\exp\left(-t/\tau^j\right)\, ,$$ where $b<1$ is a normalization
constant. Therefore, the last expression is a normalized sum and
$$\psi(t/\tau)=\tau \psi(t)/b-(1-b) \exp\left(-t/\tau\right)/b\, .$$ Using
conditions $t\gg\tau>1/b$, one obtains that at longer times
$\psi(t/\tau)=\tau \psi(t)/b$. The last expression is equivalent
to
\be\label{ex1} \psi(t)\sim 1/t^{1+\alpha}\, , \ee  %
where $\alpha=\ln(b)/\ln(1/\tau)$.

\subsection{Power law distribution of fission times}

As shown in the above example, one obtains that the pdf accounts
for all exit events from the proliferation process occurring on
all time scales, and has the power law asymptotic solution
(\ref{ex1}), where $0<\alpha<1$ and $\tau$ is a characteristic
time. Taking into account the normalization condition, one obtains
\be\label{pdf_w} %
\psi(t)= \frac{\alpha\tau^{\alpha}}{(\tau+t)^{1+\alpha}} \ee
In this case, the averaged time is infinite. A clear explanation
of Eq. (\ref{pdf_w}) can be the following quotation from Ref.
\cite{shlesinger}: ``A process with the long tailed pausing time
distribution would suffer a very sporadic behavior -- long
intermittencies may exist, followed by bursts of events. The more
probable pauses between events would be short but occasionally
very long pauses would exist. Given a long pause, there is still a
smaller but finite probability that an even longer one will occur.
It is on this basis that one would not be able to measure a mean
pausing time by examining data.''

Following statistical arguments, one should bear in mind that the
absence of a scale (or existence of a random fission scale)
follows from the variety of fluctuations of the cell environment
with which cancer cells interact \cite{Lee-Gatenby}. As stated in
\cite{garay}, the competition of proliferation and migration for
the finite free energy resources supports the mutual exclusiveness
of these cellular processes. Therefore, for the fixed cell energy,
there is a degeneracy property of many cell states that leads to
the cell state entropy\footnote{Note that the entropy is negative.
For the entropy role in carcinogenesis and tumor growth see paper
by Gatenby and Frieden \cite{Gatenby}.} $S$. As a result of this
interaction with the environment and the energy competition, as
well as of the environmental fluctuations, the escape probability
from the ``fission traps'' is described by Boltzmann's
distribution $\exp(S)$. This value is proportional to the inverse
waiting time $t^{-1}\sim\exp(S)$ \cite{bAH}. Obviously, $S$
depends on a variety of external (environmental) characteristics
and can be considered as a random value with the Poisson
distribution (for the local time and space) $P(S)=\frac{1}{\lgl
S\rgl}\exp\left(S/\lgl S\rgl\right)$, where a normalization
constant $\lgl S\rgl$ is the absolute value of the mean entropy.
The probability to find the waiting time in the interval
$(t,t+dt)$ is equal to the probability to find the entropy (as a
trapping potential) in the interval $(S,S+d(-S))$, namely,
$\psi(t)dt=P(S)d(-S)$. Taking into account these probabilistic
arguments, one obtains $\psi(t)\sim \frac{\alpha}{t^{1+\alpha}}$.
This leads to the normalized distribution of Eq. (\ref{pdf_w}),
where the transport exponent $\alpha$, as a rate of the escape, is
determined as the inverse mean entropy
\be\label{entropy}  %
\alpha=\frac{1}{\lgl S\rgl}\, .   \ee  %

\section{Kinetic equation of cell CTRW}\label{sec:kinetic}

Since the pdfs of waiting times $\psi(t)$ and jump length
$p(\mathbf{r})$ are specified in Sec.~\ref{sec:ctrw1}, we can
construct a kinetic equation of cells, following the CTRW approach
\cite{MW} and paraphrasing it from \cite{klafter,bAH}. First, we
consider a process of jumps. Let $P_j(\mathbf{r})$ be the pdf of
being at $\mathbf{r}$ after $j$ jumps. As described in
Sec.~\ref{sec:ctrw1}, the cell jumps are independent and obey the
Markov property
\be\label{markov_j} %
P_{j+1}(x)=\int P_j(\mathbf{r}')p(\mathbf{r}-\mathbf{r}')d^3r'\, ,   %
\ee %
where $P_0(\mathbf{r})$ is the initial condition.

Now, let us consider the pdf $\psi(t)$ taking into account the
dynamics of the jumps. As mentioned in Sec.~\ref{sec:ctrw1},
waiting times for different jumps are statistically independent.
Therefore, indexing the waiting time pdf by the jump number, we
define that $\psi_j(t)$ is the probability density that the $j$th
jump occurs at time $t$ (see \textit{e.g.}, \cite{bAH}, p.42).
Because of the reasonable assumption that jumps are independent
transitions, we also introduce the Markov property for
$\psi_j(t)$, which reads
\be\label{markov_w} %
\psi_{j+1}(t)=\int_0^{\infty}\psi_j(t')\psi(t-t')dt'\, ,   %
\ee  %
where $\psi_1(t)\equiv \psi(t)$. Now, we introduce the pdf
$P(\mathbf{r},t)=\sum_jP_j(\mathbf{r})\psi_j(t)$ of arriving at
coordinate $\mathbf{r}$ at time $t$. From Eqs. (\ref{markov_j})
and (\ref{markov_w}), we introduce the equation \cite{klafter}
\be\label{inteq_a}  %
P(\mathbf{r},t)=\int_{-\infty}^{\infty}p(\mathbf{r}-\mathbf{r}')
\int_0^{\infty}\psi(t-t') P(\mathbf{r}',t')d^3r'dt' +
P_0(\mathbf{r})\delta(t)\, , \ee  %
which relates the pdf $P(\mathbf{r},t)$ of just having arrived at
position $\mathbf{r}$ at time $t$ to the pdf $P(\mathbf{r}',t')$
of just arriving at $\mathbf{r}'$ at time $t'$. The last term in
Eq. (\ref{inteq_a}) is the initial condition. Thus, the pdf
$n(\mathbf{r},t)$ of being at position $\mathbf{r}$ at time $t$ is
given by arrival at $\mathbf{r}$ at time $t'$ and not moving after
this event, namely
\be\label{pdf_clP}   %
n(\mathbf{r},t)=\int_0^tP(\mathbf{r},t')\Psi(t-t')dt'\, ,   %
\ee %
where $\Psi(t)=1-\int_0^t\psi(t')dt'$ denotes the probability of
no jump during the time interval $(0,t)$. Performing the Fourier
transform $\bar{p}(k)=\hF p(x)$ and the Laplace transform
$\tilde{\psi}(s)=\hL\psi(t)$, we obtain the Montroll-Weiss
equation \cite{MW}
\be\label{MW}  %
\hat{n}(\mathbf{k},s)=\hF\hL n= \frac{1-\tilde{\psi}(s)}{s}\cdot
\frac{\hat{n}_0(\mathbf{k})}{1-\bar{p}(\mathbf{k})\tilde{\psi}(s)}\,
. \ee  %
Eq. (\ref{MW}) can be simplified for the long time $s\ll 1$ and
the large scale $k\ll 1$ asymtotics, which corresponds to the
diffusion limit $(k,s)\rightarrow(0,0)$. For the intermediate
asymptotic times, restricted from above, $t<T_{\rm max}$, we
consider the large scale limit $k\ll 1$ only. Here
$k=|\mathbf{k}|$. Taking into account the Fourier $\bar{p}(k)$
image in Eq. (\ref{MW}), where
$$p(\mathbf{r})=\frac{1}{\sqrt{4\pi\sigma_x^2}}e^{-\frac{x^2}{4\sigma_x^2}}
\frac{1}{\sqrt{4\pi\sigma_y^2}}e^{-\frac{y^2}{4\sigma_y^2}}
\frac{1}{\sqrt{4\pi\sigma_z^2}}e^{-\frac{z^2}{4\sigma_z^2}}\, , $$
one obtains
\be\label{asympt}  %
\bar{p}(k)\approx 1-\frac{\sigma^2}{6}(k_x^2+k_y^2+k_z^2)\, . \ee  %
Here, the factor $\frac{1}{6}$ is responsible for the equal
probability to jump in either direction. Substituting Eq.
(\ref{asympt}) in the Montroll-Weiss equation (\ref{MW}), one
obtains a kinetic equation in the Fourier-Laplace space.
Performing the Fourier and the Laplace inversion, one obtains the
kinetic equation of cancer cells without proliferation.
\be\label{MWasy}      %
\prt_tn(\mathbf{r},t)=
\frac{\sigma^2}{6}\Delta\int_0^t\clK(t-t')n(\mathbf{r},t')dt'\, .
\ee %
Here,  $\Delta=\prt_x^2+\prt_y^2+\prt_z^2$ and the kinetic kernel
of the transition probability
$\clK(t)\frac{s\tilde{\psi}(s)}{1-\tilde{\psi}(s)}$  is determined
from Eq. (\ref{MW}). In the next section, we consider a general
scheme of cancer cells transport provided with the antimitotic
treatment of proliferating cancer cells.

\section{CTRW analysis with antimitotic treatment}\label{sec:ctrw3}

We are concerned with the problem of the glial cancer cell
spreading in the outer invasive region as a growing tumor
spheroid, where the density of cells is $n(\mathbf{r},t)$ and
$\mathbf{r}=(x,y,z)$. Following a standard CTRW scheme for the
description of a non-Markovian transport, suggested in
Secs.~\ref{sec:ctrw1} and \ref{sec:kinetic}, we introduce a power
law waiting time probability distribution function (waiting time
pdf) $\psi(t)\sim \tau/t^{1+\alpha}$ between any two successive
cell moves/jumps, where $0<\alpha<1$ and $\tau$ is a
characteristic time scale; for example, it can be the average time
of the cell's division. This function has a general impact on both
the transition probability kernel and the proliferation kernel,
which govern the total density of cancer cells $n(\mathbf{r},t)$
in the framework of the master equation
\begin{eqnarray}\label{me}
\prt_tn(\mathbf{r},t)&=&
\frac{\sigma^2}{6}\Delta\int_0^t\clK(t-t')n(\mathbf{r},t')dt'+
\nonumber \\
&-&\int_{-\infty}^{\infty}d^3r'\int_0^t
\clC(\mathbf{r}-\mathbf{r}',t-t')n(\mathbf{r}',t')dt'\, .
\end{eqnarray}
The kinetic part of the equation is inferred in the previous
section, see Eq. (\ref{MWasy}). Here, the kinetic kernel of the
transition probability $\clK(t)$  reflects non-Markovian cell
transport and it relates to the waiting time pdf $\psi(t)$ in the
Laplace space of the Montroll-Weiss equation \cite{MW}
\footnote{In the Markov case, when
$\psi(t)=\frac{1}{\tau}e^{-t/\tau}$ and $
\clK(t-t')=\frac{1}{\tau}\delta(t-t')$, the kinetic part reduces
to the standard Fokker-Planck equation
$\prt_tn(\mathbf{r},t)=K\Delta n(\mathbf{r},t)$, where
$K=\frac{\sigma^2}{6\tau}$.}
\begin{equation}\label{clK}
\tilde{\clK}(s)=\hL[\clK(t)]=\frac{s\tilde{\psi}(s)}{1-\tilde{\psi}(s)}\,
, ~~~~~~~\tilde{\psi}(s)=\hL[\psi(t)]\, .
\end{equation}

Like any chronic antimitotic treatment, the proliferation-treating
kernel, or the TTField kernel $\clC(\mathbf{r},t)$ describes an
effect of the TTField action on the proliferating cells\footnote{
Contrary to the drug anti-mitotic action, the modality of the
TTField action is physical and related to dielectrophoresis (see,
e.g., \cite{Pohl}), when the non-uniformity of the TTField exerts
a force, focusing at the narrow cytoplasmatic bridge between two
daughter cells at fission. This leads to disruption of the
cytokinesis stage of the cell proliferation; eventually, the cells
are destroyed without the quiescent cells of normal tissues being
affected \cite{palti1,palti2,palti3}.}. In the general case, it is
a random in space and time operator. Another important property of
the cell proliferation in the outer invasive region is that this
process takes place in some fraction of the invasive volume with
some fractal dimension $\frD$. Here, the TTField action will be
considered as an averaged result of the competition between
proliferation and degradation (due to the TTField) of cells with
the maximum therapeutic effect. In this case, the dispersion of
the TTField kernel is described in the Fourier space
$C(\mathbf{k},t)=\hF[\clC(\mathbf{r},t)]$. We take this dependence
in the multiplicative form, where the TTField kernel is
proportional to the dispersion of the proliferation cell density
$C_0(\mathbf{k})$ and its probability $\Psi(t)$ to stay in the
proliferation phenotype until time $t$:
\begin{equation}\label{Psi}
C(\mathbf{k},t)=C_0(\mathbf{k})\Psi(t)=
C_0(\mathbf{k})\int_t^{\infty}\psi(t')dt'\, ,
\end{equation}
and
$\tilde{C}(\mathbf{k},s)=\hL[C(t)]=C_0(\mathbf{k})(1-\tilde{\psi}(s))/s$.
Here, the dispersion  $C_0(\mathbf{k})>0$ is also a compensation
rate between a cell's degradation and proliferation, which depends
on the density of proliferating cancer cells. At this point, we do
not specify explicitly the details of this dispersion. Note only
that in a dispersionless case, $C_0={\rm const}$, and integration
over the space in Eq. (\ref{me}) disappears. Equation (\ref{me})
in the Fourier-Laplace space reads
\begin{equation}\label{meFL}
s\hat{n}=-\frac{\sigma^2}{6}k^2\tilde{\clK}(s)\hat{n}-C_0(\mathbf{k})
\frac{1-\tilde{\psi}(s)}{s}\hat{n}+\bar{n}_0(\mathbf{k})\, ,
\end{equation}
where $\hat{n}\equiv\hat{n}(\mathbf{k},s)=\hF\hL[n(\mathbf{r},t)]$
is the Fourier-Laplace image of the pdf and $k^2=|\mathbf{k}|^2$.
Here, $\bar{n}_0(\mathbf{k})=\hF[n(\mathbf{r},t=0)]$ is the
Fourier image of the initial conditions. For simplicity, we take
it in the form of a constant value $n_0$. In this case, the pdf
$n(\mathbf{r},t)$ coincides with Green's function. Therefore, the
cell dynamics is described by the Fourier and the Laplace
inversions
\begin{equation}\label{solut}
n(\mathbf{r},t)=\hF^{-1}\hL^{-1}\left[
\frac{\bar{n}_0}{s+\frac{\sigma^2}{6}\tilde{\clK}(s)k^2+
C_0\frac{1-\tilde{\psi}(s)}{s}}\right]\, .
\end{equation}
Here and in the following, we use $C_0\equiv C_0(\mathbf{k})$. In
the Fourier space, solution (\ref{solut}) can be expressed by the
Mittag-Leffler function \cite{bateman}.  For the non-Markovian
dynamics, the Laplace image of the waiting time pdf is
$\tilde{\psi}(s)=\frac{1}{1+(\tau s)^{\alpha}}$. Therefore, the
transition probability kernel (\ref{clK}) is
$\tilde{\clK}(s)=s^{1-\alpha}/\tau^{\alpha}$, and the TTField term
is
\begin{equation}\label{clC_b}
C_0\frac{1-\tilde{\psi}(s)}{s}=
\frac{C_0\tau^{\alpha}s^{\alpha-1}}{1+(\tau s)^{\alpha}}\approx
C_{\alpha}s^{\alpha-1}\, ,
\end{equation}
where $C_{\alpha}=C_0\tau^{\alpha}$. We use $\tau s\ll 1$, which
corresponds to the consideration of the cell dynamics at the time
scale $t\gg \tau$. Substituting these expressions in Eq.
(\ref{solut}), one expands the latter over the TTField term,
defined in Eq. (\ref{clC_b}). This yields
\begin{equation}\label{solut_exp}
n(\mathbf{r},t)=n_0\hF^{-1}\hL^{-1}\left[
\sum_{l=0}^{\infty}\frac{\Big(-C_{\alpha}\Big)^l\Big(s^{\alpha-1}\Big)^{2l+1}}
{\Big(s^{\alpha}+K_{\alpha}k^2\Big)^{l+1}}\right]\, .
\end{equation}
Here, $K_{\alpha}=\frac{\sigma^2}{6\tau^{\alpha}}$ is a
generalized diffusion coefficient.

\section{The TTField kernel}

Equation (\ref{solut_exp}) can be presented in the form of the
convolution between the first term and the rest of the expansion.
The first term of the expansion corresponds to the compensated
cancer solution, when $C_0=0$. 
As presented in the previous section, the kernel $C(\mathbf{k},t)$
in Eq. (\ref{Psi}) results from the competition between
proliferation and the antimitotic TTField action. Therefore, when
$C_0=0$, the TTField compensates the proliferation of cancer
cells. We call this cancer development by ``compensated cancer
development''. Evidently, this expression makes sense only for the
cancer development in the presence of the TTField, when $C_0=0$,
and only the first term remains in expansion (\ref{solut_exp}). 
In the Fourier space, this solution can be expressed by the
Mittag-Leffler function. One obtains from Eq. (\ref{solut_exp})
\begin{equation}\label{compensat}
n_c(r,t)=\hF^{-1}\hL^{-1}\left[
\frac{n_0s^{\alpha-1}}{s^{\alpha}+K_{\alpha}k^2}\right]\, ,
\end{equation}
where the inverse Laplace transform is nothing but the definition
of the Mittag-Leffler function \cite{bateman}
\[E_{\alpha}\Big(-K_{\alpha}k^2t^{\alpha}\Big)=\int_{-i\infty}^{i\infty}
\frac{e^{st}s^{\alpha-1}ds}{s^{\alpha}+K_{\alpha}k^2}\, .\] Note
that the pdf $n_c(r,t)$ is the radial function, where
$r=|\mathbf{r}|$. Therefore, the cancer development with
proliferation compensated by the TTField is expressed by the
Fourier image of the Mittag-Leffler function
\begin{equation}\label{n_c}
n_c(r,t)=n_0\hF^{-1}
\left[E_{\alpha}\Big(-K_{\alpha}k^2t^{\alpha}\Big)\right]\, .
\end{equation}
One obtains from Eqs. (\ref{solut_exp}), (\ref{compensat}) and
(\ref{n_c})
\begin{equation}\label{solution_n}
n(\mathbf{r},t)=n_c(r,t)+n_0\hF^{-1}
\left[E_{\alpha}\Big(-K_{\alpha}k^2t^{\alpha}\Big) \star
\clR(\mathbf{k}t)\, ,\right]
\end{equation}
where the Fourier inversion is the convolution integral with the
TTField kernel
\begin{equation}\label{clR}
\clR(\mathbf{k},t)=\clR(k,t)=\hL^{-1}\left[\sum_{l=1}^{\infty}
\frac{\Big(-C_{\alpha}s^{2\alpha-2}\Big)^l}
{\Big(s^{\alpha}+K_{\alpha}k^2\Big)^l}\right]\, .
\end{equation}
Symbol $\star$ in Eq. (\ref{solution_n}) denotes the time
convolution integral.

Now, we estimate $\clR(\mathbf{k},t)$. To this end, the
denominator of the expansion should be simplified in the limit
$s^{\alpha}\gg K_{\alpha}k^2$, which is valid in the outer
invasive zone. Performing this simplification, one obtains
\begin{equation}\label{scen1}
\left[\frac{1}{\Big(s^{\alpha}+K_{\alpha}k^2\Big)^l}\right]
\approx \frac{1}{s^{\alpha l}}[1-l K_{\alpha}k^2/s^{\alpha}]\, .
\end{equation}
The Laplace inversion determines a generalization of the
Mittag-Leffler function \cite{bateman}
$E_{\alpha,\beta}(z)=\sum_{l=0}^{\infty}\frac{z^l}{\Gamma(\alpha
l+\beta)}$, where $\Gamma(\alpha l+\beta)$ (see Eq. (\ref{A10}) in
\ref{sec:app_B}); one obtains
\begin{eqnarray}\label{clR_1}
&\clR(k,t)=-C_{\alpha}t^{1-\alpha}
E_{2-\alpha,2-\alpha}\Big(-C_{\alpha}t^{2-\alpha}\Big) +
\nonumber \\
&+K_{\alpha}k^2C_{\alpha}t\sum_{l=0}^{\infty}
\frac{(l+1)(-1)^lC_{\alpha}^lt^{(2-\alpha)l}}{\Gamma[(2-\alpha)l+2]}
\equiv \clR_1+\clR_2\, .
\end{eqnarray}
Let us first treat the first term in Eq. (\ref{clR_1}). The
argument of the Mittag-Leffler functions is large, which yields
the power-law behavior \cite{klafter,bateman}
\begin{equation}\label{MLF}
E_{\alpha,\beta}(z)\sim\frac{-1}{\Gamma(\beta-\alpha)z}\, .
\end{equation}
In this case, the TTField kernel needs to be treated in the
framework of the generalized functions (distributions)
\cite{mainardi}, which yields $\clR_1=\delta(t)$ for the kernel in
the convolution integral. This contribution to the distribution
function in the Fourier space is the inversion of the
Mittag-Leffler function, which compensates $n_c(r,t) $ exactly:
\begin{equation}\label{solut_fin}
n_c(r,t)-n_0\hF^{-1}
\left[E_{\alpha}\Big(-K_{\alpha}k^2t^{\alpha}\Big)\right]=0\, .
\end{equation}

Therefore, the solution in Eq. (\ref{solution_n}) is determined by
the convolution with the second term $\clR_2$ of the TTField
kernel in Eq. (\ref{clR_1}). To obtain the analytical expression,
some simplification of the gamma function should be performed.
This reads
\begin{equation}\label{simple_1}
\frac{l+1}{\Gamma[(2-\alpha)l+2]}=
\frac{l+1}{[(2-\alpha)l+2]\Gamma[(2-\alpha)l+1]}  
\approx \frac{1}{2\Gamma[(2-\alpha)l+1]}\, .
\end{equation}
Again, we arrive at the two-parameter Mittag-Leffler function,
whose asymptotic behavior (\ref{MLF}) yields the kernel
$\clR_2=\frac{\alpha K_{\alpha}k^2}{2\Gamma(\alpha)}t^{\alpha-1}$.
Therefore, one obtains the fractional integration of the
Mittag-Leffler function with the TTF kernel
\begin{equation}\label{FracInt_1}
n(r,t)=\frac{n_0\alpha K_{\alpha}}{2} 
\hF^{-1}\left[\frac{k^2}{\Gamma(\alpha)}
\int_0^t(t-t')^{\alpha-1}
E_{\alpha}\Big(-K_{\alpha}k^2{t'}^{\alpha}\Big)dt'\right]\, .
\end{equation}
Note that $n(r,t)$ is the radial function. The integration over
the time is well defined and yields (see \textit{e.g.},
\cite{podlubny} [Eq. (1.100)])
\begin{equation}\label{FracInt_1a}
n(r,t)=\frac{n_0\alpha K_{\alpha}}{2} \hF^{-1}\left[ t^{\alpha}k^2
E_{\alpha,1+\alpha}\Big(-K_{\alpha}k^2t^{\alpha}\Big) \right]\, .
\end{equation}
In the outer invasive zone for the initial time scenario, when
$K_{\alpha}k^2t^{\alpha}< 1$ is valid, the Mittag-Leffler function
constitutes the stretched exponential behavior
\cite{klafter,bateman}
\[E_{\alpha,1+\alpha}\Big(-K_{\alpha}k^2t^{\alpha}\Big)\sim
\exp[-\lambda K_{\alpha}k^2t^{\alpha}]\, ,\] where
$\lambda=\Gamma(1+\alpha)$. Note also that we work with the
three-dimensional Fourier transform. Therefore, since
$k^2=k_x^2+k_y^2+k_z^2$, differentiating over $\lambda$ reduces
the problem to the simple Fourier inversion of the Gaussian
packets, which yields
\begin{eqnarray}\label{solut_sc1}
n(r,t)&=&-\frac{\prt}{\prt\lambda}\frac{n_0\alpha}{2\sqrt{(4\pi\lambda
K_{\alpha} t^{\alpha})^3}} \exp\left[-\frac{r^2}{4\lambda
K_{\alpha}t^{\alpha}}\right] \nonumber  \\
&\propto&
r^2(K_{\alpha}t)^{-\frac{3\alpha}{2}}\exp\Big(-\frac{r^2}{K_{\alpha}t^{\alpha}}\Big)\,
.
\end{eqnarray}
This solution (\ref{solut_sc1}) exhibits the result of the TTField
action on the glioma development that leads to subdiffusion, or
the stretched exponential restriction of the cancer cells
spreading. An important fact here is that expression
(\ref{solut_sc1}) is independent of $C_0=C_0(\mathbf{k})$ and,
correspondingly, is independent of any dispersive properties of
the tissues interacting with the TTField.

As observed, the TTField cannot stop the cancer spreading as the
averaged radius of the outer invasive zone increases
subdiffusively $\lgl r^2(t)\rgl\sim t^{\alpha}$. An example of
compensated cancer invasion with time is shown in Fig.~1. This
weakening of the TTField action is due to the
migration-proliferation dichotomy, and is reflected in the
convolution treatment term with the power-law time delay kernel
$C(t)=C_0(\mathbf{k})\Psi(t)$ in Eq. (\ref{me}). The final result
in Eq. (\ref{solut_sc1}) is independent of the treatment rate as
well.
\begin{figure}[htbp]
\includegraphics[width=0.9\hsize]{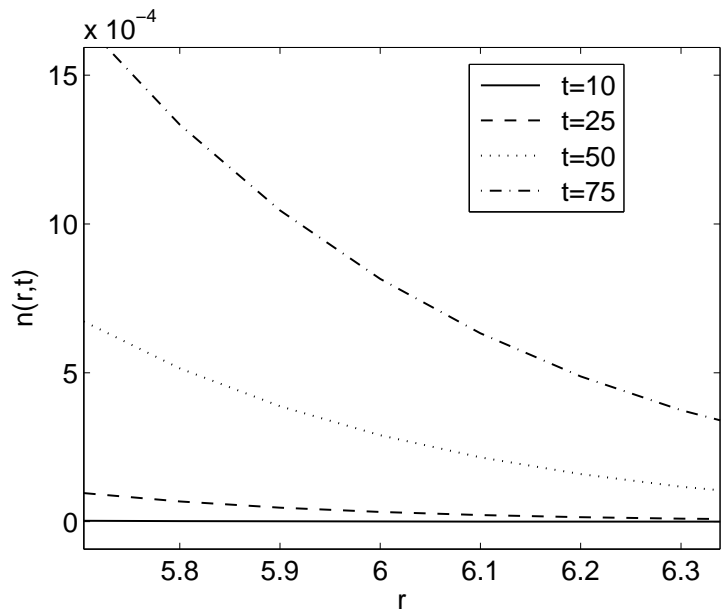}
\caption{Increasing concentration of cancer cells with time for
the dimensionless radius $r$ and time, according to Eq.
(\ref{solut_sc1}). Different plots correspond to different times:
from bottom to top $t=10,25,50,75$. Here, $K_{\alpha}=1$ and
$\alpha=1/3$. In the vicinity of $r=6$, the condition of the outer
invasive zone $K_{\alpha}t^{\alpha}/r^2\ll 1$ is fulfilled.}
\label{fig:draw1}
\end{figure}

\section{Untreated cancer development}\label{sec:untreat}

The TTField action, as any other antimitotic treatment, is
important and leads to an increase in the survival period as a
result of the essential decrease in the cancer spread rate. To
understand this phenomenon, one needs to compare the spread of
cancer cells with and without the TTFields. For example, let us
consider the overall velocity of the constant concentration front
propagation in the case of treated and untreated cancer. It
follows from Eq. (\ref{solut_sc1}) that the front of constant
concentration corresponds, approximately, to the solution $r^2\sim
\frac{\alpha}{2} K_{\alpha}t^{\alpha}\ln t$. Therefore, for
$C_0=0$, the spread rate of the treated cancer decreases with time
$v_0=\dot{r}\sim K_{\alpha}t^{\frac{\alpha}{2}-1}$ and vanishes at
the asymptotics $t\rightarrow\infty$. The situation changes
dramatically for untreated cancer. In the present analysis, it
corresponds to $C_0<0$, where in the absence of the TTField
$|C_0|$ is now the rate of proliferation of cancer cells. Cell
proliferation contributes strongly to the transport, since the
density of cancer cells $n(r,t)$ increases with time. It is well
known that this process leads to the nonzero velocity of the front
propagation \cite{petrovskii,murray,fir2011}, which has the form
$v_C\sim \sqrt{C_0K_{\alpha}}$. Therefore, the TTField leads to
the failure of the asymptotic front propagation of the cancer
spread, while the front of untreated cancer spreads with a
constant velocity.

Another important property is the mean squared displacements (MSD)
$\lgl r^2(t)\rgl$, which can be calculated for both treated and
untreated cancers. There are various scenarios of the cancer
spread and some approaches have been presented in
\cite{Khain,Deutsch,Chauviere,kolobov,fir2011}. Here, it is
instructive to consider the cancer spread in the presence of the
migration-proliferation dichotomy as a renewal process (see
\textit{e.g.} \cite{cox,fellerII}), following \cite{fir2011}. The
switching between migrating and proliferating phenotypes for
untreated cancer leads to the superdiffusive/sub-ballistic spread
of cancer cells, when the transport exponent is larger than 1.

From the compensated cancer solution (\ref{solut_sc1}), one
obtains that the MSD for the treated cancer reads
\begin{equation}\label{untreat_1}
\lgl r^2(t)\rgl\sim K_{\alpha}t^{\alpha}\, ,
\end{equation}
which corresponds to subdiffusion. For untreated cancers, a
reversible process of switching between migration and
proliferation phenotypes leads to the increase in the cancer cell
population. Therefore, to apply the renewal theory with the
constant cell population, we consider the dynamics of cells with
the migration-proliferation switching that are simultaneously
moving together with the front with the constant velocity $v_C$.
This yields the following scheme. The random cell transport is
considered for the constant number of cancer cells, while
proliferation contributes to advection with the constant velocity
of the cancer cells $v_C$ along the radial direction.

Now we take into account the process of migration-proliferation
switching. The residence time pdf for the cell proliferation,
described by Eqs. (\ref{ex1}) and (\ref{pdf_w}), is $\psi(t)\sim
(\tau/t)^{1+\alpha}$ with the infinite averaged residence time.
Contrary to the proliferation time, the averaged transport time
$\lgl t\rgl=\bar{t}$ is finite. During this time, cancer cells
move the distance $r(\bar{t})=v_c\bar{t}$. To find the MSD $\lgl
r^2(t)\rgl$ for the asymptotically large time, we consider that
the displacement of a cell at time $t$ is proportional to the
number of migration-proliferation switchings $N(t)$, such that $
r(t)\sim v_C\bar{t}N(t)$. Therefore, one obtains for the MSD $\lgl
r^2(t)\rgl\ge {\lgl r(t)\rgl}^2 \sim [v_C\bar{t}]^2{\lgl
N(t)\rgl}^2$.

Note that the number of cells is constant, and therefore, the
distribution function $n(r,t)$ of cancer cells to be at the
position $r(t)$ at time $t$ can be considered a function of the
number of switchings $n(r,t)\equiv p(\clN,t)={\rm Pr}[N(t)=\clN]$.
Using a well known result from the renewal theory (see
\textit{e.g.} \cite{cox,fellerII}), one obtains that the Laplace
transform of $p(\clN,t)$ is determined by $\tilde{\psi}(s)$
\begin{equation}\label{untreat_2}
\tilde{p}(\clN,s)=\frac{\tilde{\psi}^{\clN}[1-\tilde{\psi}]}{s}\,
.
\end{equation}
One obtains the second moment $\lgl
N^2(t)\rgl=\sum_{\clN=0}^{\infty}\clN^2p(\clN,t)$, which reads in
the Laplace space
\begin{equation}\label{untreat_3}
\lgl
\tilde{N^2}(s)\rgl=\sum_{\clN=0}^{\infty}\clN^2\tilde{p}(\clN,s)
=\frac{1-\tilde{\psi}}{s}\sum_{\clN=0}^{\infty}\clN^2\tilde{\psi}^{\clN}\,
.
\end{equation}
Taking into account that for the large time asymptotic
$\tilde{\psi}(s)\approx 1-(s\tau)^{\alpha}$, one obtains after the
Laplace inversion that $\lgl
N^2(t)\rgl\sim(t/\tau)^{2\alpha}/\Gamma(1+2\alpha)$. Finally, one
obtains that the MSD reads
\begin{equation}\label{untreat_4}
\lgl
r^2(t)\rgl\sim\frac{[v_c\bar{t}]^2}{\Gamma(1+2\alpha)}
\left(\frac{t}{\tau}\right)^{2\alpha}\,
.
\end{equation}
As follows from Eqs. (\ref{untreat_1}) and (\ref{untreat_4}), the
main benefit of any antimitotic treatment, including the TTField,
is decreasing the transport exponent of the cancer spread twofold.
Moreover, the changes can be more dramatic, when
$\frac{1}{2}<\alpha<1$. In this case the difference between
treated and untreated cancer corresponds to the difference between
a subdiffusive cancer spread and a superdiffusive one.


\section{Conclusion}\label{sec:discuss}

Understanding of glioma cancer kinetics in the presence of the
tumor treating field (TTField) can be important for the further
development of the TTField's efficiency for medical treatment. It
should be stressed that the TTField technology is already an
officially accepted method for cancer treatment\footnote{See
http://www.novocure.com/}. We have shown that glioma development
under the TTField treatment can be modelled in the framework of
the continuous time random walk, where the TTField segregation of
proliferating from not proliferating cells is naturally accounted.
The constructed reaction-transport equation (\ref{me}) is based on
the fact that the cytokinesis affects the cell dynamics strongly,
and the cell kinetics is non-Markovian, which is reflected in both
the transport and reaction terms of Eq. (\ref{me}). As was stated
in Ref. \cite{garay} ``As a matter of fact, in one single cell,
cytokinesis and migration are separated temporally; but on the
level of a cell population - and this is the case of tumors - cell
migration and proliferation occurs simultaneously.'' Obviously,
for the CTRW consideration, which accounts for the dynamics of the
entire population at the individual particle level, the dynamics
of one single cell with the migration-proliferation dichotomy is
the most important.

The TTField technology, eventually, fails to stop glioma cancer
development. This conclusion also correlates with the recent
empirical (clinical) data, observed in \cite{Turner}, where the
effect of the field strength on glioblastoma multiforme response
was studied in patients treated with the TTField. The authors
observed that cancer cells respond to the TTFields in such a way
that they avoid the field affect. To some extent, this is indirect
clinical evidence that supports the hypothesis that the main
reason for the treatment failure is the migration-proliferation
dichotomy, which is completely independent of the field. It should
be noted that TTFields belong to a wide class of therapies that
are effective against the abnormal proliferation of transformed
cells. However, as compared to chemotherapy regimens, the
treatment modality is completely different, being of a physical
nature, and accompanied by minimal local side effect and no
systemic side effects \cite{palti3,Turner}. Therefore, therapy
efficiency can be improved by combining chemotherapy with the
TTField, as was stated in \cite{Turner}.


The kernel $C({\bf r},t)$ describes the competition between
proliferation and degradation due to the TTField action. This
complicated function is random in space and time, and the result
depends essentially on its sign. When it is negative, which means
that the treatment is sufficient, or is absent,  the kernel
(\ref{clR}) is no longer the TTField kernel,  and corresponds to
proliferation. Since the argument in the Mittag-Leffler function
is positive, this leads to the exponential growth of the density
of cancer cells $\sim\exp(C_{\alpha} t)$. In this case, the
nonlinear mechanism becomes important in that it restricts the
unlimited growth of cell numbers, see \textit{e.g.}
\cite{petrovskii,murray}.

First, an analytical attempt to understand the TTField's influence
on glioma development has been suggested by virtue of the
geometrical construction of both the anomalous cell transport and
the inhomogeneous distribution of proliferating cells in the
framework of a fractional comb model for the 1D \cite{iom2012} and
the 3D \cite{iom2013} analysis. It has been shown there that the
efficiency of the medical treatment depends essentially on the
mass fractal dimension of the cancer in the outer invasive zone.
In the present research, a generalization of these geometrical
constructions of the TTField action was suggested. The solution
for the cell density of the developing glioma  $n({\bf r},t)$ was
obtained for any arbitrary distribution of proliferating cells.
Moreover, the obtained final result in Eq. (\ref{solut_sc1})
corresponds to the \textit{compensated} cancer solution with
$C({\bf r},t)=0$. In this case, any inhomogeneities do not affect
subdiffusion of cancer cells that leads to subdiffusion of free
particles. This result is valid for any arbitrary $C({\bf r},t)$.
To some extent, one can anticipate this property of the
intermediate asymptotic result (\ref{solut_sc1}), that all
proliferating cells disappear, and correspondingly, the TTField
becomes ineffective. As a result, a clone of migrating cancer
cells spreads freely according to Eq. (\ref{solut_sc1}) without
the details of the degradation rate of proliferating cells due to
the TTField having an influence. In this case, the TTField plays
the role of an experimental/technological separation mechanism for
the surviving migrating cells only. Probably, this process can be
important for understanding the mechanism of the
migration-proliferation dichotomy.

One should recognize that the obtained results correspond to the
intermediate asymptotic solution according to Eq. (\ref{scen1}),
when time is restricted from above $t<T_{\rm max}\sim
\Big(\frac{r_{\rm
max}^2}{K_{\alpha}}\Big)^{\frac{1}{\alpha}}=\Big(\frac{r_{\rm
max}^2}{\sigma^2}\Big)^{\frac{1}{\alpha}}\tau$, where $T_{\rm
max}/\tau\gg 1$. Taking $T_{\rm max}$ as the maximum survival time
of the order of $10^3$ days \cite{palti3} and $1/\tau \sim 1 {\rm
(day)}^{-1}$ as the averaged proliferation rate \cite{garay}, one
obtains $T_{\rm max}/\tau \sim 10^3$, which corresponds to the
intermediate asymptotics approximation, performed in Eq.
(\ref{scen1}). Therefore, the main conditions for the performed
approximations that are valid for the intermediate asymptotic
times $t$ in the outer-invasive region $r$ are $1\gg
(\tau/t)^{\alpha}\gg \sigma^2/r^2$. Taking into account that the
outer-invasive region is of the order of $r\sim 1~{\rm cm}$ and
$\sigma\sim 10^{-3}~{\rm cm}$, one obtain that these inequalities
are valid and verify the obtained result.

Finally, it is tempting to find what happens if a patient has a
possibility to survive beyond the time $T_{\rm max}$, when
$t\rightarrow\infty$. In this case, when $s\rightarrow 0$, the
Tauberian theorem \cite{fellerII}, applied to Eq. (\ref{clR}),
gives $R(k,t)\sim\delta(t)$, which yields
$n(r,t\rightarrow\infty)\equiv 0$. Again, one anticipates this
result at the infinite time scale, when \textit{all} migrating
cells have a possibility to be converted into proliferating cells
with further degradation due to the TTField. Therefore, one has to
consider a stationary value problem (for $t\rightarrow\infty$)
with a constant flux of cancer cells from the boundary of the
tumor core, and the result with $n(r,t\rightarrow\infty)\equiv 0$
is no longer valid. Moreover, since the initial value problem of
Eq. (\ref{me}) is described by the fractional Fokker-Planck
equation, the problem of a stationary solution should be revised
carefully. This problem will be the subject of future studies.

I thank E. Yodim for language editing. This work was supported by
the Israel Science Foundation (ISF).

\appendix

\section{Fractional integro--differentiation}\label{sec:app_B}

\def\theequation{A. \arabic{equation}}
\setcounter{equation}{0}

The consideration of a non-Markovian process in the framework of
kinetic equations leads to the study of the so-called fractional
Fokker-Planck equation, where time processes are not local
\cite{klafter}. In this case, time derivations are substituted by
time integration with the power law kernels. One arrives at
so-called fractional integro--differentiation.

A basic introduction to fractional calculus can be found, {\em
e.g.}, in Ref. \cite{podlubny,SKM}. Fractional integration of the
order of $\alpha$ is defined by the operator
\be\label{A1}  %
{}_aI_t^{\alpha}f(t)=\frac{1}{\Gamma(\alpha)}
\int_a^tf(\tau)(t-\tau)^{\alpha-1}d\tau, ~~(\alpha>0)\, , \ee %
where $\Gamma(\alpha)$ is a gamma function. There is no constraint
on the limit $a$. In our consideration, $a=0$ since this is a
natural limit for the time. A fractional derivative is defined as
an inverse operator to ${}_aI_t^{\alpha}\equiv I_t^{\alpha}$ as
$\frac{d^{\alpha}}{dt^{\alpha}}=I_t^{-\alpha}=D_t^{\alpha}$;
correspondingly $ I_t^{\alpha}=\frac{d^{-\alpha}}{dt^{-\alpha}}
=D_t^{-\alpha}$. Its explicit form is convolution
\be\label{A2}  %
D_t^{\alpha}=\frac{1}{\Gamma(-\alpha)}\int_0^t
\frac{f(\tau)}{(t-\tau)^{\alpha+1}}d\tau \, . \ee    %
For arbitrary $\alpha>0$, this integral is, in general, divergent.
As a regularization of the divergent integral, the following two
alternative definitions for  $D_t^{\alpha} $ exist \cite{mainardi}
\be\label{A3} %
{}_{RL}D_{(0,t)}^{\alpha}f(t)\equiv D_{RL}^{\alpha}f(t)=
D^nI^{n-\alpha}f(t) 
\frac{1}{\Gamma(n-\alpha)}\frac{d^n}{dt^n}\int_0^t
\frac{f(\tau)d\tau}{(t-\tau)^{\alpha+1-n}} \, , \ee %
\be\label{A4} %
D_C^{\alpha}f(t)=
I^{n-\alpha}D^nf(t)  
\frac{1}{\Gamma(n-\alpha)}\int_0^t
\frac{f^{(n)d\tau}(\tau)}{(t-\tau)^{\alpha+1-n}} \, , \ee  %
where $ n-1<\alpha<n,~~n=1,2,\dots$. Eq. (\ref{A3}) is the
Riemann--Liouville derivative, while Eq. (\ref{A4}) is the
fractional derivative in the Caputo form \cite{podlubny,mainardi}.
Performing integration by part in Eq. (\ref{A3}) and then applying
Leibniz's rule for the derivative of an integral and repeating
this procedure $n$ times, we obtain
\be\label{A5} %
D_{RL}^{\alpha}f(t)=D_C^{\alpha}f(t)+\sum_{k=0}^{n-1}f^{(k)}(0^+)
\frac{t^{k-\alpha}}{\Gamma(k-\alpha+1)} \, . \ee   %
The Laplace transform can be obtained for Eq. (\ref{A4}). If
$\hL[f(t)]=\tilde{f}(s)$, then
\be\label{A6}  %
\hL\left[D_C^{\alpha}f(t)\right]=s^{\alpha}\tilde{f}(s)-
\sum_{k=0}^{n-1}f^{(k)}(0^+)s^{\alpha-1-k}\, . \ee   %
The following fractional derivatives are helpful for the present
analysis
\be \label{A8}
D_{RL}^{\alpha}[1]=\frac{t^{-\alpha}}{\Gamma(1-\alpha)}\, , ~~
D_C^{\alpha}[1]=0\, . \ee %
We also note that
\be\label{A9}
D_{RL}^{\alpha}t^{\beta}=\frac{t^{\beta-\alpha}\Gamma(\beta+1)}
{\Gamma(\beta+1-\alpha)}\, , \ee  %
where $\beta>-1$ and $\alpha>0$.
The fractional derivative from an exponential function can be
simply calculated as well by virtue of the Mittag--Leffler
function (see {\em e.g.}, \cite{podlubny,bateman}):
\be\label{A10}   %
E_{\gamma,\delta}(z)=\sum_{k=0}^{\infty}
\frac{z^k}{\Gamma(\gamma k+\delta)} \, . \ee   %
Therefore, we have the following expression
\be\label{A11}   %
D_{RL}^{\alpha}e^{\lambda t}=t^{-\alpha}E_{1,1-\alpha}(\lambda
t)\, . \ee

\end{document}